\documentclass[]{aastex631}

\usepackage{amsmath}

\begin{document}

\title{Radiation Signatures of Electron Acceleration in the Decelerating Jet of MAXI J1348-630}

\author{Aishwarya Sarath}
\affiliation{St Joseph's University \\
Langford Road , Bangalore \\ Karnataka,India}

\author[0000-0002-8434-5692]{Markus B\"ottcher}
\affiliation{Center for Space Research \\North-West University \\
Potchefstroom,South Africa }

\begin{abstract}
A discrete jet component (blob) ejection and its subsequent deceleration were observed in the 2019/2020 outburst of the low-mass X-ray binary MAXI J1348–630. A first kinematic analysis of the deceleration due to an abrupt transition from an evacuated cavity to the interstellar medium suggested a kinetic energy exceeding $10^{46}$ erg, surpassing estimates of the available total ejection energy. However, incorporating a transition layer with exponential density growth between the cavity and the interstellar medium recently enabled a kinematic analysis with much more realistic energy requirements of approximately $10^{44}$ erg. 
Here, we study the expected radiative signatures of electrons accelerated within the decelerating blob by introducing a model akin to the relativistic blast wave model for gamma-ray bursts, considering radiative energy losses and radiation drag, to simulate the deceleration of a relativistically moving plasmoid. This model yields snap-shot spectral energy distributions and multi-wavelength light curves from synchrotron and synchron-self-compton (SSC) emission. Notably, the synchrotron emission peaks in the X-rays, but the predicted X-ray flux is negligible compared to thermal emission from the accretion disk. The predicted radio light curve closely resembles the observed one during the jet decleration phase following the outburst in 2019/2020.

\end{abstract}

%% Keywords should appear after the \end{abstract} command. 
%% The AAS Journals now uses Unified Astronomy Thesaurus concepts:
%% https://astrothesaurus.org
%% You will be asked to selected these concepts during the submission process
%% but this old "keyword" functionality is maintained in case authors want
%% to include these concepts in their preprints.

\keywords{{\href{https://astrothesaurus.org/uat/33}{Astrophysical jets (33)}} ; {\href{https://astrothesaurus.org/uat/1907}{X-ray binaries (1907)}}}

\section{Introduction} \label{sec:intro}

MAXI J1348-630 is classified as a black-hole low-mass X-ray binary (BH LMXB) system, where a low-mass companion star fills its Roche lobe, feeding a hot accretion disk surrounding the compact object. In a subset of X-ray binary systems, including MAXI J1348-630, a fraction of the accreted matter is channeled into mildly relativistic jets, in which case they are termed microquasars. They are regarded as scaled-down counterparts of radio-loud active galactic nuclei (AGNs), sharing many of the same features. This resemblance provides a valuable opportunity to study the scale-invariant properties of black holes, especially those related to the interplay between accretion and ejection, as these processes evolve through various regimes and timescales \citep{Kording,Carotenuto_2021,10.1093/mnrasl/slac093}.

An effective approach for identifying X-Ray Binaries (XRBs) and studying their state changes involves continuous X-ray monitoring of the entire sky. This is particularly useful because most XRBs are transient sources and display unpredictable X-ray outbursts \citep[e.g.,][]{2020ApJ...899L..20T,2000A&A...359..251C,hannikainen1998radiomonitoringgx3394,2003A&A...397..645M}. The MAXI (Monitor of the All-Sky X-Ray Image) instrument, mounted on the International Space Station, scans about 85~\% of the sky every 92 minutes \citep{Matsuoka2009,Koljonen_2019}. It captures each section of the sky for 40 -- 100 seconds during each orbit. MAXI is specifically designed for detecting and studying black hole X-ray binary systems. Since 2009, MAXI has found 13 new XRBs and regularly observed their state changes.

Synchrotron emission from compact jets is responsible for the radio to infrared emission observed in black-hole X-ray binares \citep[BHXRBs; ][]{fender2003jets,2019MNRAS.488L..18K}. Black hole transients remain in a quiescent state for extended periods, often lasting several months to years or even decades. During their outbursts, they emit radiation across a wide range of wavelengths, from radio waves to X-rays \citep[e.g.,][]{Carotenuto21}. During an outburst, the system transitions through various accretion states, each with distinct spectral and timing characteristics \citep{ Remilard, Belloni2010, Belloni2016}. Initially, it exhibits a rising hard X-ray state dominated by non-thermal power-law emission due to inverse-Compton scattering by hot electrons near the compact object. As the accretion rate increases, the system moves into an intermediate state where the X-ray spectrum softens due to increased thermal emission from the accretion disk \citep{Corbel_2003,2013ApJ...775....9H,Chen_2019,markoff2003exploring}. Eventually, it reaches the soft state, characterized by dominant thermal emission from an optically thick, geometrically thin accretion disk truncating at the innermost stable circular orbit (ISCO). Later, the system returns to a hard state before slowly approaching quiescence \citep{  Ingram, ZdZMarek, Poutanen2014, Steiner}.

Some BH LMXBs launch transient jets, which are relativistic plasma blobs ejected in opposite directions \citep{Mirabel, Hjellming}. During outbursts, the jets of BH LMXBs evolve significantly. In the hard state, compact jets emit self-absorbed synchrotron radiation, resulting in a flat or slightly inverted radio spectrum \citep{Corbel2000, Fender2001}. These jets can contribute to hard X-ray emission and may accelerate matter to relativistic bulk Lorentz factors of a few \citep{Markoff, ribo2004asymmetriccompactjetgrs}. In the soft state, the jets are quenched, with a significant drop in radio emission \citep{Fender1999, Corbel2000}. When transitioning back to the hard state, jets are gradually reactivated \citep{MillerJones, Kalemci}. These jets can display apparent superluminal motion and emit optically thin radio waves as they interact with the interstellar medium \citep{Corbel, Tomsick_2003}. Although transient jets are associated with strong radio flares near state transitions, the exact jet-launching mechanism remains unclear \citep{Fender2004, Corbel2004}.

MAXI J1348-630 was discovered by the MAXI detector during a giant outburst on 2019 January 26 \citep[MJD 58,509,][]{2019ATel12425....1Y}. Using MeerKAT and Australia Telescope Compact Array (ATCA) observations, \cite{Carotenuto21} observed the ejection of compact radio jet components associated with the major X-ray outburst at speeds $\ge 0.69$~c, and their subsequent deceleration. Different estimates  of the distance to the source exist: approximately $2.2^{+0.5}_{-0.6}$ kpc using H I absorption  \citep{10.1093/mnrasl/slaa195}, and about $3.4 \pm 0.3$ kpc based on X-ray observations of a dust-scattering halo \citep{2021A&A...647A...7L}. In this work, we use the value of $D = 2.2$ kpc.

The {\it Neil Gehrels Swift} X-Ray Telescope (XRT) monitored the source  with 85 pointings over a period of almost 1 year, beginning shortly after the discovery of the outburst on January 26, 2019 \cite[see][for more information on the observations and data reduction]{Carotenuto_2021}. As discussed in \cite{Carotenuto_2021}, the outburst began with a rapid increase in X-ray brightness, which was first detected on MJD 58509. Initially, the system was in the hard state, characterized by a power-law X-ray spectrum indicative of emission from a hot corona. As the outburst progressed, the X-ray emission became softer, transitioning through different states: from the hard state to the intermediate state (IMS) on February 3, 2019 (MJD 58517), and then to the soft state on February 8, 2019 (MJD 58522.6). During the soft state, the X-ray spectrum was dominated by thermal emission from the accretion disk, with a peak disk temperature of around 0.8~keV. The system remained in the soft state as it reached peak flux of the outburst and then began a gradual decay. After about two months, the system transitioned back to the IMS on March 26, 2019 (MJD 58597), and eventually returned to the hard state on April 2, 2019 (MJD 58604). The optical counterpart of MAXI J1348-630 was discovered on 2019 January 26 using the iTelescope.Net T31 instrument in Siding Spring, Australia. A new optical transient at R.A. = 13h48m12.88s, Decl. = -63$^\circ$16'28.4" (J2000.0) with a magnitude of 16.18 was identified within the Swift error circle, 82 arcseconds from its center.
\citep{,2019ATel12430....1D,2019ATel12434....1K,2019ATel12448....1N,russell2019optical}

The initial, major outburst was followed by three smaller flares over the course of the next few months, labelled R1 -- R3, in accordance with the nomenclature of \cite{Carotenuto_2021}. The slow decay of the X-ray light curve during the initial outburst is well described by disk-dominated emission. As described in \cite{Carotenuto_2021}, during this phase, the strong jets might have cleared out or displaced the surrounding ISM, creating a low-density cavity around the black hole system. The presence of this cavity is the basis for the cavity -- ISM transition models of \cite{Carotenuto22} and \cite{Zdziarski_2023} for the radio-jet deceleration. 

The deceleration of the compact jet components ejected by MAXI J1348-630 in tandem with the  radio flare RK1, was modelled by \cite{Carotenuto22}, assuming a sharp transition from an evacuated cavity to  the interstellar medium (ISM). Their kinematic analysis required the ejecta to have an initial energy of $E_0 = 4.6^{+20}_{-3.4} \times 10^{46} \, (\phi/1^o) (n_{\rm ISM}/(1 \, {\rm cm}^{-3}))$~erg, where $\phi$ is the opening angle of the jet and $n_{\rm ISM}$ the particle density of the ISM. This energy is significantly beyond the limit corresponding to magnetically arrested accretion onto a maximally rotating stellar-mass black hole. As a sharp transition from an evacuated cavity to the ISM is likely not a realistic assumption, \cite{Zdziarski_2023} introduced an extension of the model of \cite{Carotenuto22}, considering a transition layer between the cavity and the ISM with exponential density growth. The kinematic analysis using such a density profile resulted in a much more realistic kinetic energy value of about $\sim 10^{44}$~erg.

In this paper, we study the multi-wavelength radiative signatures expected from the acceleration of particles at the shock front formed by the interaction of the ejected plasmoid ('blob') with the surrounding medium. We describe our analysis of the kinematics of the decelerated blob, the resulting relativistic electron distribution, and the evaluation of snap-shot SEDs and radio light curve fits in \S \ref{sec:Model}. Numerical results are presented and compared to observations in \S \ref{sec:results}. We summarize and discuss our results in \S \ref{sec:discussion}. 

\section{Model of a Decelerating Jet} \label{sec:Model}

Following \cite{Zdziarski_2023}, we assume a thin transition layer between the evacuated cavity (with density $n_{\rm cavity} \ll 1$~cm$^{-3}$) and the ISM with standard density $n_{\rm ISM} = 1$~cm$^{-3}$. The inner boundary of the transition layer is at a distance $r_c$ from the black hole, from where the density begins to grow exponentially with an e-folding distance $dr$ until it reaches the density of the ISM at $r_{ISM}$:

\begin{equation} \label{eq:1}
n(r) =
\begin{cases}
    n_{\text{cavity}} & \text{if } r \leq r_{\text{c}}, \\
    n_{\text{cavity}} \, e^{(r - r_{\text{c}})/dr} & \text{if } r_{\text{c}} \leq r \leq r_{\text{ISM}}, \\
    n_{\text{ISM}} & \text{if } r \geq r_{\text{ISM}}.
\end{cases}
\end{equation}

\begin{equation}\label{eq:2}
r_{\text{ISM}} = r_{\text{c}} + d_r \ln q, \qquad q = \frac{n_{\text{ISM}}}{n_{\text{cavity}}}
\end{equation}

In the next section we will discuss the kinamatics of a mildly relativistic plasmoid interacting with an external medium with the above density profile.

\subsection{\label{sec:Jet kinematics}Jet kinematics}

To model the kinematics of a decelerating blob using a momentum-conservation approach, we use the methods of \cite{BP09}, which were developed to model a quasi-exponential light-curve decay observed in the blazar 3C279. We here develop a scaled-down version of that model appropriate for the scales of microquasars. The work of \cite{BP09}, in turn, was based on an adaptation of the blast wave model for GRB afterglows \citep[e.g.][]{Chiang_1999} to blazar jets.
A plasmoid moving with initial mass \(M_0\) and initial bulk Lorentz factor $\Gamma_0$ moves along the jet. As it moves, it sweeps up mass from the external medium, thus increasing its mass $M$. The total momentum P of the plasmoid in the stationary frame is $P = \beta \Gamma M c$ and is conserved, except for momentum imparted on anisotropically emitted radiation. If the plasmoid emits energy evenly in all directions in its rest frame, as is commonly assumed for synchrotron and synchrotron-self-Compton (SSC) emission, its momentum remains constant over time, $dP/dt = 0$. However, if a significant amount of particle energy is lost to Compton scattering of an external radiation field (EC = External Compton), the momentum in this EC flux is lost to the plasmoid. Hence, the equation governing the rate of change of momentum, $(dP/dt)$, can be expressed as \citep{BP09}

\begin{equation}
\label{eq:3}
\frac{dP}{dt} = \left(\frac{dP}{dt}\right)_{\text{EC}} = \frac{c M \dot{\Gamma}}{\beta} + \Gamma \beta \dot{M} c.
\end{equation}

Although in the specific model setup for MAXI J1348-630, EC radiation is negligible compared to synchrotron and SSC emission, our code does account for EC energy and momentum losses, using Eqs. (3) and (4) in \cite{BP09}, in order to keep the model more generally applicable.

Relativistic mass-energy in the plasmoid is not only gained by sweeping up external material, but also lost through radiative dissipation. Hence, the change in relativistic mass may be expressed as

\begin{equation}
\label{eq:5}
\dot{M} = A(r)\rho(r)\Gamma(r) \frac{dr}{dt} + \frac{1}{c^2} \dot{E}_{\text{rad}},
\end{equation}

The radial velocity $dr/dt$ is given by $\beta c$ with $\beta = \sqrt{1 - 1/\Gamma^2}$. The cross-section of the jet $A(r)$ is given by $\pi R_b^2(r)$ with $R_b(r)$ being the radius of the jet. $\rho(r)$ represents the density of the external material being swept up by the plasmoid. The radiative energy loss is given by 

\begin{equation}
\label{eq:6}
\dot{E}_{\text{rad}} = -\frac{1}{\Gamma} \frac{4}{3} c \sigma_T u \int_{\gamma_{\text{min}}}^{\gamma_{\text{max}}} {N_e(\gamma)} \gamma^2 \, d\gamma
\end{equation}
where \(u\) is the sum of the energy densities in the magnetic and radiation fields $(u_B + u_{\text{ext}} + u_{\text{sy}})$, and we assume that Compton scattering is occurring predominantly in the Thomson regime.

Derivatives with respect to time can be converted into derivatives with respect to  the distance $r$ from the central engine using the radial velocity above. This leaves us with the set of coupled differential equations for $\Gamma(r)$, $M(r)$, and $\beta(r)$:

\begin{equation}\label{eq:7}
\frac{d\Gamma}{dr} = -\Gamma(r)\beta^2(r) \frac{dM/dr}{M(r)} + \frac{\Gamma^2 \dot{E}_{\text{EC}}}{4\pi Mc^3}
\end{equation}
\begin{equation}\label{eq:8}
\frac{dM}{dr} = A(r) \rho(r) \Gamma(r) + \frac{\dot{E}_{\text{rad}}}{\Gamma(r)\beta(r) c^3}
\end{equation}
\begin{equation} \label{eq:81}
\frac{d\beta}{dr} = {\frac{d\Gamma}{dr}}{\frac{1}{\beta \Gamma^3}}
\end{equation}

As in \cite{BP09}, the radiating relativistic electron energy distributions is obtained assuming a continuous injection of relativistic particles instantaneously accelerated into a power-law spectrum, described by an injection function \( Q(\gamma) = Q_0 \gamma^{-q} H(\gamma_{\text{min}}, \gamma, \gamma_{\text{max}}) \). The low-energy cutoff, \(\gamma_{\text{min}}\), is determined by the fraction of swept-up power transferred to relativistic electrons, while the high-energy cutoff, \(\gamma_{\text{max}}\), is set by balancing the acceleration timescale with the synchrotron loss timescale. The critical electron energy, \(\gamma_c\), is is the energy beyond which particles radiate away energy faster than they can escape, leading to a broken power law distribution depending on whether the cooling is slow or fast. In the slow-cooling regime, \(\gamma_c > \gamma_{\text{min}}\), and in the fast-cooling regime, \(\gamma_c < \gamma_{\text{min}}\). This results in the broken power-law distributions as used in \cite{BP09},  depending on whether the system is in the slow-cooling or the fast-cooling regime:

\begin{equation} \label{eq:z}
N_e(\gamma) = N_0
\begin{cases}
    \left( \frac{\gamma}{\gamma_b} \right)^{-p_1} & \text{if } \gamma_1 < \gamma < \gamma_b \cr
\left( \frac{\gamma}{\gamma_b} \right)^{-p_2} & \text{if } \gamma_b < \gamma < \gamma_2
\end{cases}
\end{equation}

Here, \( N_0 \) is the normalization constant. In the slow cooling regime, the parameters \( \gamma_1 \), \( \gamma_b \), and \( \gamma_2 \) correspond to \( \gamma_{\text{min}} \), \( \gamma_c \), and \( \gamma_{\text{max}} \), respectively. The indices \( p_1 \) and \( p_2 \) correspond to \( q \) and \( q + 1 \). In the fast cooling regime the parameters \( \gamma_1 \), \( \gamma_b \), and \( \gamma_2 \) correspond to $\gamma_c$, $\gamma_{\text{min}}$, and $\gamma_{\text{max}}$, respectively. The indices \( p_1 \) and \( p_2 \) correspond to \( 2 \) and \( q + 1 \). In the case of MAXI J1348-630, we find that the electron energy distribution is always in the slow cooling regime.

\subsection{Analytic solutions to the equation of motion}\label{sec:analytical}

Eqs.  \ref{eq:7} and \ref{eq:8} constitute the equations of motion derived from momentum conservation. If the jet is conical, then $A(r) = \Omega_j r^2$, where $\Omega_j$ is the solid angle subtended by the jet; if it is cylindrical, then $A(r) \equiv A_0$.
Analytical solutions to the coupled differential equations \ref{eq:7} and \ref{eq:8} may be found in the limiting case where the swept-up mass \(M_{\text{sw}}\) is negligibly small compared to the initial mass \(M_0\), i.e., \(M_0 \gg M_{\text{sw}}\), and the radiative energy and momentum loss terms are negligible, hence keeping only the first terms on the r.h.s. of Eqs. \ref{eq:7} and \ref{eq:8}. The former condition limits the applicability of the solution to the early slowdown phase, either with \(\rho(r) = \rho_c\) for \(r \leq z_c\) or \(\rho(r) = \rho_c e^{(r - z_c)/dz}\) for \(z_c \leq r \leq z_c + dz\).

\subsubsection{Motion inside the cavity}

Inside the cavity ($r < z_c$), the equation of motion for $\Gamma(r)$ for the conical-jet case reduces to
\begin{equation}\label{eq:32}
\frac{d\Gamma(r)}{dr} \approx -\frac{\Gamma^2(r) \beta^2(r) \Omega_j \rho_c}{M_0 } r^2= -\frac{(\Gamma^2(r) - 1) \Omega_j \rho_c}{M_0 }r^2
\end{equation}
Separation of variables and integration yields an analytical solution of the form
\begin{equation}\label{eq:51}
\ln \left(\frac{\Gamma(r) + 1}{\Gamma(r) - 1}\right) - \ln \left(\frac{\Gamma_0 + 1}{\Gamma_0 - 1}\right) = \frac{2}{3} \Omega_j \rho_c \frac{r^3 - r_0^3}{M_0} \equiv B_{\rm con}(r)
\end{equation}
In the case of a cylindrical jet, the solution for \(r < z_c\) simplifies to
\begin{equation}\label{eq:15}
\ln \left(\frac{\Gamma(r) + 1}{\Gamma(r) - 1}\right) - \ln \left(\frac{\Gamma_0 + 1}{\Gamma_0 - 1}\right) = \frac{2A_0 \rho_c}{M_0} (r - r_0) \equiv B_{\rm cyl}(r).
\end{equation}
Defining,
\begin{equation}\label{eq:17}
C_{\rm con/cyl}(r) \equiv \frac{\Gamma_0 + 1}{\Gamma_0 - 1} e^{B_{\rm con/cyl}(r)}
\end{equation}
one finds the final analytical solution as
\begin{equation}\label{eq:18}
\Gamma(r) = \frac{C_{\rm con/cyl}(r) + 1}{C_{\rm con/syl}(r) - 1}
\end{equation}
One easily verifies that for $r = r_0$, $\Gamma(r) = \Gamma_0$ and for $r \rightarrow \infty$, $\Gamma(r) \rightarrow 1$, as expected.

\subsubsection{Motion inside the transition zone}

In the transition zone ($z_c < r < z_c + dz$) Eq. \ref{eq:7} for a conical jet reduces to

\begin{equation}\label{eq:19}
\frac{d\Gamma(r)}{dr} \approx -\frac{(\Gamma^2(r) - 1) \Omega_j \rho_c e^{-\frac{z_c}{d_z}} r^2 e^{\frac{r}{d_z}}}{M_0 }
\end{equation}
Again, separation of variables and integration yields
with $\Gamma_c = \Gamma(r_c)$. 
%{\bf \textcolor{red}{Please quote the solution for the conical jet here.}}
\begin{equation}\label{eq:solution_conical}
\ln \left(\frac{\Gamma(r) + 1}{\Gamma(r) - 1}\right) - \ln \left(\frac{\Gamma_c + 1}{\Gamma_c - 1}\right) = \frac{2 \Omega_j \rho_c e^{-\frac{z_c}{d_z}}}{M_0} \left(\frac{r^3 e^{\frac{r}{d_z}}}{3} - \frac{r_c^3 e^{\frac{r_c}{d_z}}}{3}\right).
\end{equation}

In the case of a cylindrical jet, we have

\begin{equation}\label{eq:dGdr_conical}
\frac{d\Gamma(r)}{dr} \approx -\frac{(\Gamma^2(r) - 1) A_0 \rho_c e^{-\frac{z_c}{d_z}} e^{\frac{r}{d_z}}}{M_0 }
\end{equation}
with the  solution  
\begin{equation}\label{eq:20}
\ln \left(\frac{\Gamma(r) + 1}{\Gamma(r) - 1}\right) - \ln \left(\frac{\Gamma_c + 1}{\Gamma_c - 1}\right) = \frac{2A_0 \rho_c e^{-\frac{z_c}{d_z}}}{ M_0}  \left(e^{\frac{r}{d_z}} - e^{\frac{r_c}{dz}}\right)  d_z
\end{equation}
These equations may be solved for \(\Gamma(r)\) analogous to Eq. \ref{eq:18}. One again easily sees
that \(\Gamma(r_c) = \Gamma_c\) and \(\Gamma(r \to \infty) \to 1\).

\begin{figure}[htbp]
    \centering
    \includegraphics[width=0.5\textwidth]{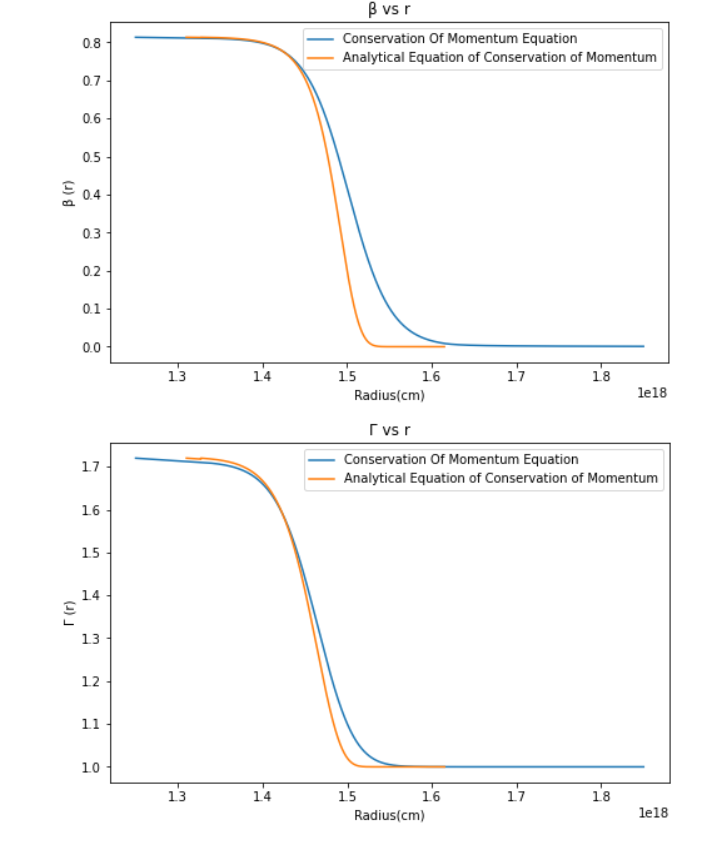}
    \caption{Comparison of the analytical and numerical solutions to the plasmoid deceleration process in a cylindrical jet.}
    \label{fig:deceleration_comparison}
\end{figure} 

For our numerical modeling of MAXI J1348-63, we considered a cylindrical jet shape. Fig. \ref{fig:deceleration_comparison} shows the comparison between the full numerical solutions for $\Gamma(r)$ and $\beta(r)$ and these analytical approximations. The figure illustrates that the agreement is good while the blob is still inside the cavity and entering the transition zone, but as more and more mass is swept up and radiative energy losses become substantial, the approximations break down, as expected. Therefore, for the purpose of our numerical study of the radiative signatures of this model, we used the full numerical solutions. Yet, this comparison to analytical solutions instills confidence that our numerical solutions describe the deceleration process accurately.

\begin{table}[ht]
    \centering
    \begin{tabular}{|c|c|c|}
        \hline
        \textbf{Parameter} & \textbf{Symbol} & \textbf{Value} \\
        \hline
        Initial Lorentz Factor & $\Gamma_0$ & 1.70 \\
        \hline
        Initial Mass & $M_0$ & $4.3 \times 10^{23}$ g \\
        \hline
        Plasmoid Radius & $R_{b}$ & $2.3 \times 10^{16}$ cm \\
        \hline
        Relativistic Electron Fraction & $\xi_{e}$ & 0.5 \\
        \hline
        Transition Region & $r_c$ & $1.327 \times 10^{18}$ cm \\
        \hline
        Interstellar Medium Density & $n_{\text{ISM}}$ & $1 \, \text{cm}^{-3}$ \\
        \hline
        Cavity Density & $n_{\text{Cavity}}$ & $10^{-5} \, \text{cm}^{-3}$ \\
        \hline
        Magnetic Field Strength & $B$ & 0.0043 G \\
        \hline
        Electron Injection Index & $q$ & 2.5 \\
        \hline
        efolding distance & $d_r$ & $2.5 \times 10^{16}$ cm \\
        \hline
    \end{tabular}
    \caption{Parameters for the plasmoid evolution model for MAXI~J1348-630}
    \label{tab:Parameters}
\end{table}

\subsection{\label{sec:radiation}Radiative Output}

We used the \texttt{JetSet} \citep{jetset3,jetset2,jetset1} software to calculate the synchrotron and synchrotron self-Compton (SSC) spectral energy distributions (SEDs). The computation of the synchrotron SED in \texttt{JetSet} follows the approach outlined by \cite{dermer2010highenergyradiationblack} and \cite{2008ApJ...686..181F} which utilize the approximation provided by \cite{Aharonian2010} for the pitch-angle-averaged synchrotron power, which simplifies the integration over the time-evolving electron distribution.

For the SSC SED, \texttt{JetSet} again follows the methodologies described by \cite{dermer2010highenergyradiationblack} and \cite{2008ApJ...686..181F}. The SSC process involves the scattering of synchrotron photons by the same electrons that emitted them.  
The Compton scattering kernel used \citep{PhysRev.167.1159} accounts for the kinematic constraints and cross section for Compton scattering in both the Thomson and Klein-Nishina regimes.

We can estimate the external radiation energy density due to direct  accretion-disk photons as $u_{\text ext} \sim L_d / (4 \pi r^{2}) \sim 10^{-16} \text{erg cm}^{-3}$ which is much smaller than the magnetic-field energy density, $u_B \sim 10^{-7} \text{erg cm}^{-3}$. To our  knowledge, no measurement of the IR - optical -  UV emission from the putative companion star exists in the literature, suggesting that the corresponding radiation energy density is also negligible. Therefore, we have omitted EC radiation.

\begin{figure}[htbp]
\gridline{\fig{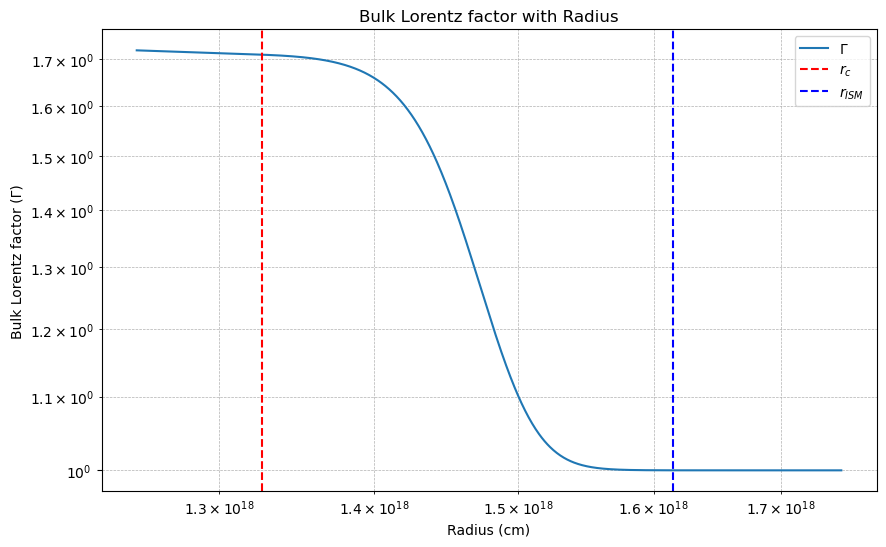}{0.6\textwidth}{(a)}}
\gridline{\fig{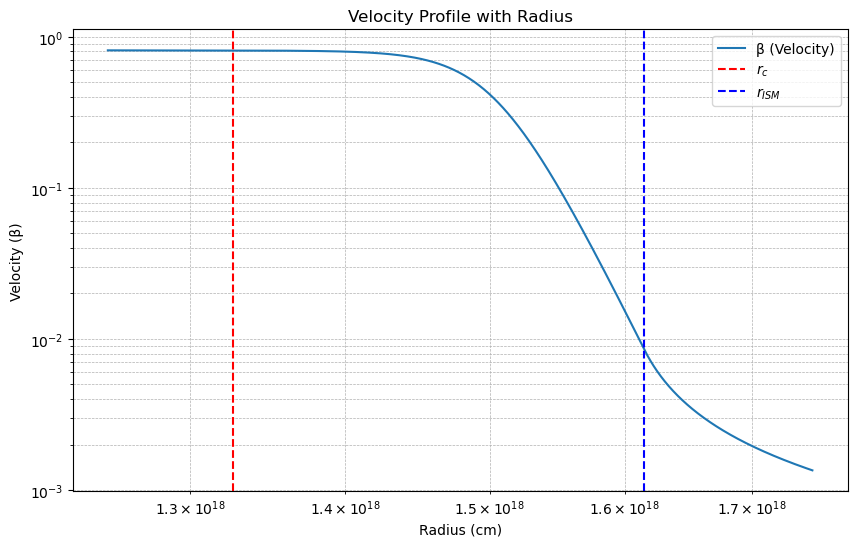}{0.6\textwidth}{(b)}}
\gridline{\fig{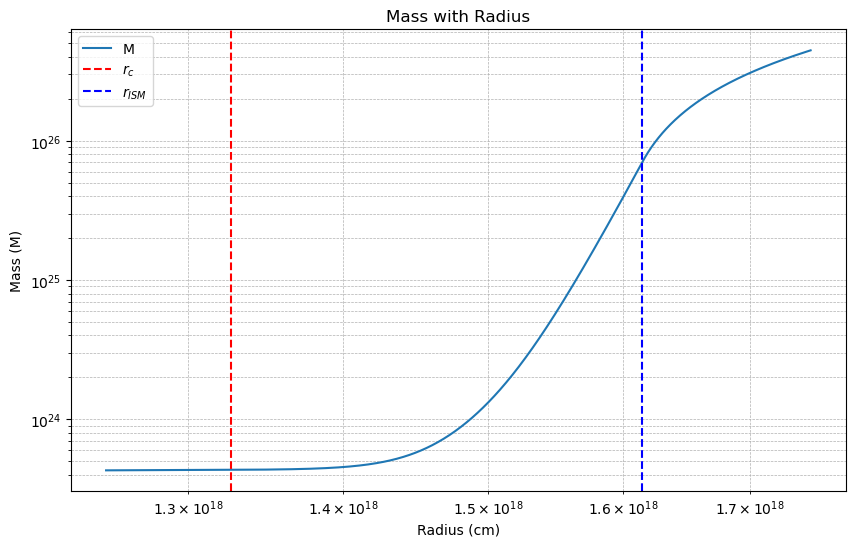}{0.6\textwidth}{(c)}}
\caption{(a) Bulk Lorentz factor $\Gamma$. (b) Dimensionless velocity $\beta$. (c) Total mass of the plasmoid $M$. The dashed vertical lines indicate $r_c$ (red) and $r_{\rm ISM}$ (blue).}
\label{fig:combined}
\end{figure}

\section{\label{sec:results}Numerical results: Spectral energy distributions (SEDs) and multi-wavelength light curves}

The decelerating-blob model outlined above was applied specifically to the observed plasmoid deceleration of the radio jet component RK1 in MAXI J1348-630 in 2020. The jet geometry was assumed to be cylindrical. The density profile (Eq.  \ref{eq:1}), initial bulk Lorentz factor, and other parameters, as listed in Table \ref{tab:Parameters}, are taken or derived from \cite{Zdziarski_2023}. The jet viewing angle $\theta$ was kept at 30$^{\circ}$. 

{Figure \ref{fig:combined} illustrates the evolution of the bulk Lorentz factor $\Gamma$, the normalized velocity  $\beta$, and  total mass of the plasmoid with distance, obtained by numerically solving the coupled set of equations \ref{eq:7}, \ref{eq:8}, and \ref{eq:81}. As the plasmoid approaches the transition to the ISM, the bulk Lorentz factor levels off at 1, while the the dimensioneless velocity continues to decrease. As in the case of GRBs, we see that the plasmoid evolves through a coasting phase where the bulk Lorentz factor remains approximately constant at $\Gamma_0$ and begins to decelerate around the charecteristic deceleration radius $r_d$, where the relativistic swept-up mass equals the initial mass of the plasmoid \citep[e.g.,][]{CB98,Dermer99}.

\begin{figure}
    \centering
    \includegraphics[width=0.9\linewidth]{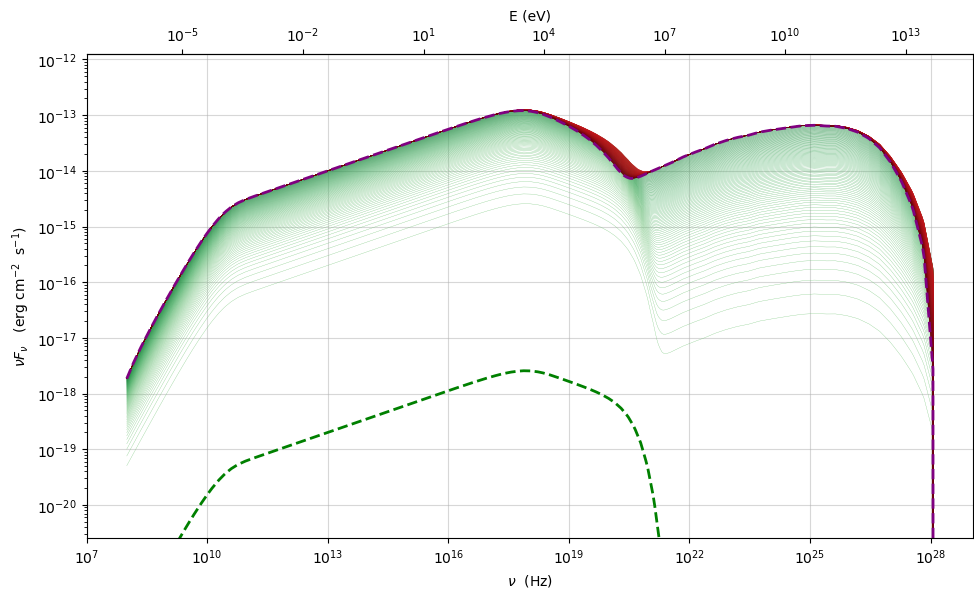}
    \caption{Temporally evolving SED, starting from green dotted lines via red solid lines to purple dotted lines}
    \label{fig:fig3}
\end{figure}

{Figure \ref{fig:fig3}  presents the temporally evolving SEDs from radio to $\gamma$ rays from a simulation that provides a fit to the decay of the radio light curve of the decelerating component RK1 observed in MAXI J1348-630 (see below). Notably, the synchrotron emission peaks in the X-ray band,  while the SSC spectrum peaks at multi-GeV -- TeV $\gamma$-ray energies, suggesting that such sources may be detectable by the Fermi Large Area Telescope (LAT) or ground-based Cherenkov Telescope facilities. There is a very small shift in the evolving SED peaks over time, thus predicting no significant time delays between the light curves at different frequencies. 

\begin{figure}[ht]
    \centering
    \includegraphics[width=0.95\textwidth]{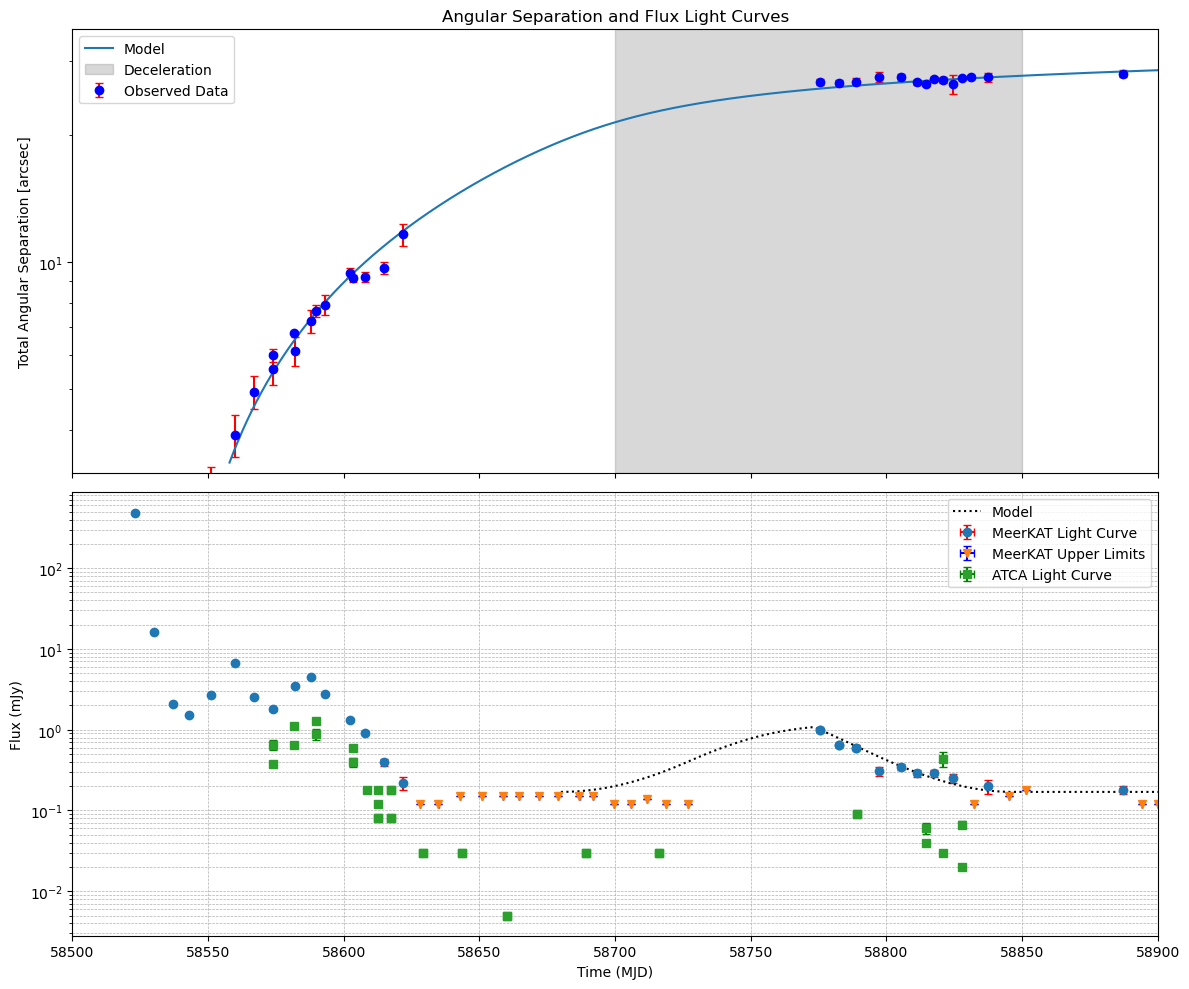}
    
    \caption{Fit of the decelerating-plasmoid model to the MeerKAT and ATCA radio light curve \citep[data from][]{Carotenuto_2021} the decelerating component RK1 (bottom panel), and of the angular seperation of RK1 vs. time (top panel).}
    \label{fig:fit}
\end{figure}

The bottom panel of Figure \ref{fig:fit} shows the fit of our model to the angular separation of component RK1 from the central black hole of MAXI J1348-630 as a function of time, as already demonstrated by \cite{Zdziarski_2023}. The bottom panel  of Figure \ref{fig:fit} shows the radio light curve of RK1 along with the prediction of our model at a frequency of 1.3~GHz. It illustrates that the decay phase of the radio light curve is well modelled, but the model over-predicts the flux around MJD 58700 -- MJD 58730, which could be due to the upper limits from MeerKAT . The maximum 1 -- 10~keV X-ray flux predicted by the model, $\sim 10^{-13}$~erg~s$^{-1}$~cm$^{-2}$, falls significantly below the observed Swift-XRT flux. This suggests that the X-ray flux is dominated by thermal contributions from the accretion disk and corona, and the jet emission is negligible.

\section{\label{sec:discussion}Summary and discussion}

In this paper, we modelled the radio light curve of the low-mass X-ray binary system MAXI~J1348-630 during an episode of deceleration of a radio component of the jet of this system. \cite{Zdziarski_2023} has shown that a smooth transition between an evacuated cavity around the binary and the ISM, with which the ejected jet component interacts, is capable of reproducing the observed jet deceleration with realistic energy requirements.  
Motivated by this development, we {investigate the radiative signatures of a decelerating plasmoid in the jet of the MAXI J1348-630 microquasar which was first detected in early February 2019, around MJD ~58520, corresponding to the beginning of the first outburst phase \citep{Russell_2019} . This model incorporates the inertia of swept-up material as the plasmoid travels through the surrounding medium, alongside the effects of radiation drag and cooling. 

Our analysis reveals that, akin to the relativistic blast-wave model used for gamma-ray bursts (GRBs) \cite{1999ApJ...512..699C}, the plasmoid undergoes a transition from an initial coasting phase with a nearly constant Lorentz factor to a self-similar deceleration phase. Our model successfully captures the characteristics of the RK1 radio lightcurve that coincided with the observed radio-component deceleration phase, lasting approximately one week. Our analysis predicts no significant time delays between the emissions in different wavelength bands. Our model SEDs (see section \ref{sec:radiation}) exhibit a synchrotron peak around X-ray wavelengths, 
whose flux, however, remains below the dominant thermal contribution from the accretion disk}, and an SSC peak at multi-GeV -- TeV $\gamma$-ray energies, suggesting that such systems may also be detectable by space- or ground-based $\gamma$-ray observatories. 

A very similar decelerating-blob model had previously been used to obtain a fit to a quasi-exponentially decaying optical lightcurve of the blazar 3C279 \citep{BP09}. It offers a promising way to interpret multi-wavelength data of other microquasars in outburst, in which jet ejections and subsequent radio-jet component decelerations are observed. Most notably, XTE J1550–564 showed such an episode around the year 2000. This microquasar was spatially resolved in both radio and X-ray wavelengths, with the radio emission confirmed to be from synchrotron processes, while the origin of the X-rays is still under debate \citep{Corbel_2002}. Proper motion observations using Chandra showed that the eastern jet decelerated between 1998 and 2000, and within about two years, the western jet experienced significant deceleration as well. Also in this case, a smooth transition between an evacuated cavity and the ISM might substantially reduce the energy requirements of the jet ejection \citep{Zdziarski_2023}. The X-ray light curve of the eastern jet had an exponential decay with a \(e\)-folding time of \(t_{\text{decay}} \sim (338 \pm 14)\) days, which is consistent within uncertainties with the western jet (\(t_{\text{decay}} \sim [311 \pm 37]\) days). At the same time, the radio light curve showed a much steeper decay compared to the X-ray light curve \citep{Tomsick2003,M3}. In future work, we will apply our decelerating-blob model to this system to evaluate whether it is able to reproduce the observed kinematics of the radio and X-ray jet components and the corresponding light curves.

\section{Acknowledgments}
The authors thank the anonymous referee for constructive criticism and suggestions that have helped to substantially improve the manuscript, and Prof. A. A. Zdziarski for fruitful discussions.
The work of MB is supported by the South African Department of Science and Innovation and the National Research Foundation through the South African Gamma-Ray Astronomy Program (SA-GAMMA).

\bibliography{sample631}{}
\bibliographystyle{aasjournal}

\end{document}